\documentclass[journal]{vgtc}              

\onlineid{1772}

\vgtccategory{Research}

\title{Charting Public Health: A Taxonomic Study of Visualization Practices in the Public Health Field}

\author{%
  Mia Hines
  and 
  Alvitta Ottley
}

\authorfooter{
  \item
  	Mia Hines is with Washington University in St. Louis
  	E-mail: hines.mia@wustl.edu
  \item
  	Alvitta Ottley is with Washington University in St. Louis
  	E-mail: alvitta@wustl.edu
}

\abstract{%
  Public health organizations regularly produce and publish data visualizations to raise awareness of critical issues, influence decision-making processes, and promote overall well-being. However, the design practices shaping these visualizations in real-world settings remain largely unexamined, limiting the research community's ability to evaluate their effectiveness, accessibility, and alignment with communication goals. To address this gap, we construct and analyze a large-scale corpus of over 4,000 real-world data visualizations drawn from more than two dozen websites associated with U.S. and international public health organizations. We evaluate salient design characteristics like chart type, visualization accessibility, use of embellishments like iconography, and design flaws. This work contributes to understanding real-world decisions in designing data visualizations and supports public health officials in improving data visualization-related communications. Visualizations in our finalized corpus and the labeled dataset can be found at \url{https://washuvis.github.io/ChartingPublicHealth/}.
  
}


\teaser{
  \centering
  \includegraphics[width=\linewidth, alt={Figure 1 displays the pipeline for identifying stable design profiles from the collected public health organizations' visualizations. The five design profiles are tagged as Statistical, Narrative, Geospatial, Domain-Focused, and Biomedical.}]{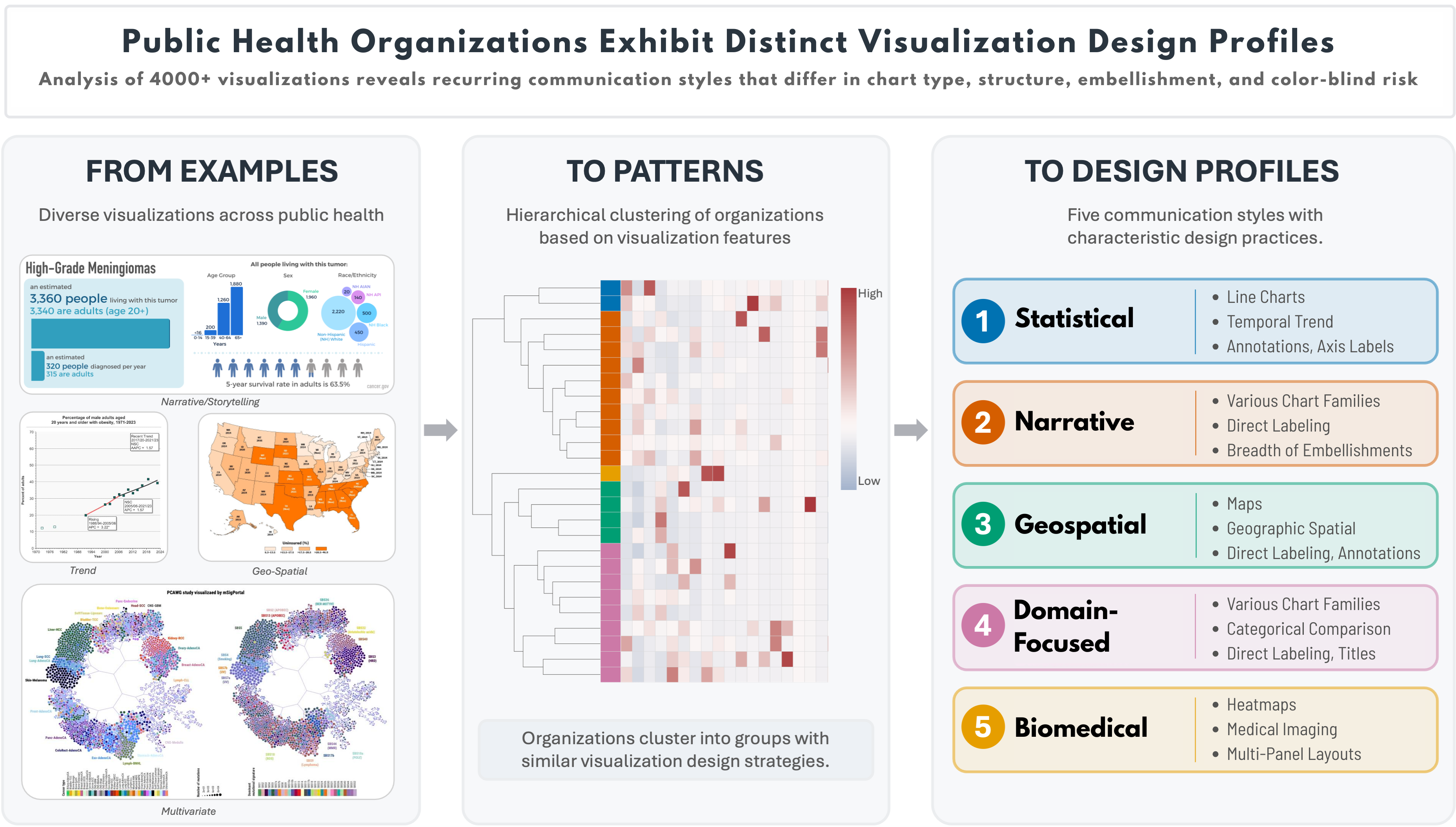}
  \caption{%
  	We have clustered organizations across a large corpus of real-world visualizations and uncovered five recurring communication styles that differ in chart forms, structural completeness, annotation practices, and visual embellishments. The resulting profiles reveal systematic differences between evidence-forward reporting, public storytelling, scientific dissemination, and imaging-heavy communication.%
  }
  \label{fig:teaser}
}

\graphicspath{{figs/}{figures/}{pictures/}{images/}{./}} 

\usepackage{tabu}                      
\usepackage{booktabs}                  
\usepackage{lipsum}                    
\usepackage{mwe}                       
\usepackage{ccicons}                   
\usepackage{soul}
\usepackage[dvipsnames]{xcolor}
\usepackage{tabularx}
\usepackage{amsmath}
\usepackage{listings}
\usepackage{dblfloatfix}

\usepackage{adjustbox}
\usepackage{tikz}
\usetikzlibrary{arrows.meta, positioning, shapes.geometric, calc, fit}

\usepackage{mathptmx}                  

\usepackage{array}
\usepackage{tabularx}

\newcolumntype{C}[1]{>{\centering\arraybackslash}m{#1}}
\newcolumntype{Y}[1]{>{\raggedright\arraybackslash}m{#1}}

\definecolor{MyNote}{HTML}{3b158c}
\definecolor{accent}{RGB}{37, 99, 235}
\definecolor{clusterOne}{HTML}{0072B2}   
\definecolor{clusterTwo}{HTML}{D55E00} 
\definecolor{clusterThree}{HTML}{009E73}     
\definecolor{clusterFour}{HTML}{CC79A7}    
\definecolor{clusterFive}{HTML}{E69F00}       

\sethlcolor{red!20}

\newcommand{\Nnonprofit}{18~}
\newcommand{\Ninternational}{9~}
\newcommand{\NUS}{30~}

\newcommand{\acro}[1]{\textsc{\MakeLowercase{#1}}}

\newcommand{\statistical}{{\color{clusterOne}{\acro{\textbf{Statistical}}}}}
\newcommand{\narrative}{{\color{clusterTwo}{\acro{\textbf{Narrative}}}}}
\newcommand{\geospatial}{{\color{clusterThree}{\acro{\textbf{Geospatial}}}}}
\newcommand{\domain}{{\color{clusterFour}{\acro{\textbf{Domain-Focused}}}}}
\newcommand{\biomedical}{{\color{clusterFive}{\acro{\textbf{Biomedical}}}}}

\newcommand{\stepX}[1]{{\small\circnum[#1]{#1}~}}
\newcommand{\circnum}[2][]{%
  \tikz[baseline=(n.base)]\node[
    draw,
    circle,
    inner sep=0.45ex,
    minimum size=3.6ex,
    line width=0.8pt,
    fill=blue!40,
    draw=blue!40,
    text=white
  ] (n) {\normalsize \bfseries #2};}
\begin{document}


\firstsection{Introduction}

\maketitle


Data visualizations are among the most consequential communication artifacts in public decision-making systems. In domains such as public health, they communicate evidence, shape policy discourse, and influence population behavior.  These visualizations appear in a wide range of public-facing materials, including reports, dashboards, infographics, and web content, and play a central role in shaping how health information is understood by policymakers, practitioners, and the general public. For instance, visualizations are consequential for monitoring disease outbreaks, e.g., COVID-19 and HIV/AIDS, and dispatching resources to communities in need \cite{baxter_development_2022, sullivan2020data, legenza_geospatial_2023}. Interactive visualizations are used for electronic health records (EHRs), thereby reducing the cognitive load on medical staff and improving in-patient care \cite{lee_public_2015, senathirajah_characterizing_2020}. Visualization-focused web applications have been developed to increase shareholder engagement and improve policy-makers' understanding of critical issues \cite{lammons_involving_2023, park_understanding_2021, zakkar_interactive_2017}. Thus, understanding the design decisions of public health visualizations has direct implications for evaluating comprehension, trust, and action in high-stakes settings ~\cite{pandey2023you,mckinley2025trustworthy}. 

Despite their importance, there is limited empirical understanding of how data visualizations are actually designed and deployed in real-world public health settings. Prior work has largely focused on the development of visualization applications \cite{sullivan2020data, cocoros2021riskscape}, literature surveys of specific design factors \cite{ofori2025visual, park2022impact}, or has examined practitioner needs through interviews \cite{park_understanding_2021}. While these efforts provide valuable insights into what \emph{could} be effective, they offer limited visibility into what is \emph{actually produced at scale}. As a result, fundamental questions remain unanswered: How do organizations communicate data in practice? What design choices dominate real-world deployment? And what tradeoffs or vulnerabilities do these choices introduce?

In this work, we address this gap through a large-scale empirical characterization of real-world visualization practice in the public health domain. We construct and analyze a corpus of more than 4,000 visualizations collected from over two dozen U.S. and international public health organizations. Using a semi-automated pipeline that combines web scraping, computational filtering, and manual validation, we capture a broad, ecologically grounded sample of visualizations as they are encountered by real audiences.

 Our analysis reveals a highly concentrated real-world design space. A small number of chart families dominate the corpus, with line charts alone accounting for over 40\% of all visualizations and temporal trend communication emerging as the dominant task context. While organizations generally adhere to core structural conventions, we find that important vulnerabilities persist, including inconsistent use of legends and direct labels and widespread color accessibility risks. Notably, nearly one-quarter (23.08\%) of visualizations contain potentially risky color pairs under simulated color-vision-deficiency conditions.

Beyond corpus-level trends, we identify distinct \emph{organization-level design profiles}, revealing that institutions exhibit stable, interpretable visualization styles. These styles are likely shaped by subject matter, communication norms, audience expectations, and reporting goals. 
Although grounded in public health, these findings help expose broader patterns in how organizations operationalize visualization design under real-world constraints and make the following contributions:

\begin{itemize}[noitemsep]
    \item A large-scale, ecologically grounded corpus and annotation framework for studying real-world visualization practice in public health, and provides reusable infrastructure for future research.
    
    \item A corpus-level empirical characterization of dominant chart forms, communication tasks, structural conventions, and accessibility vulnerabilities.
    
    \item An organization-level analysis revealing stable design profiles that expose how institutional norms shape visualization practice.
\end{itemize}

   \begin{table*}[tb]
      \caption{%
      	Public Health organizations within the final corpus. 
      }
      \label{tab:org_counts}
      \scriptsize%
      \newlength{\digitwidth}%
      \settowidth{\digitwidth}{0}%
      \setlength{\tabcolsep}{12pt}
      \centering%
      \begin{tabu}{%
      	  l%
      	  	*{2}{@{\hspace{1pt}}l@{\hspace{1pt}}}%
            *{2}{@{\hspace{5pt}}c@{\hspace{5pt}}}%
      	}
      	\toprule
      	Organization Name & Acronym & Type & Images Collected & Final Visualizations \\
      	\midrule
          	Agency for Toxic Substances and Disease Registry & ATSDR & National Agency &  923 & 97  \\
          	Alzheimer's Association & AA & Nonprofit Organization &   1,118 &  10 \\
          	American Cancer Society & ACS & Nonprofit Organization &  546 &  29   \\
          	Centers for Medicare \& Medicaid Services & CM\&MS & National Agency &  360 &  33  \\
            Dana-Farber Cancer Institute & DCI & Nonprofit Organization & 1,018 & 27 \\
            Fogarty International Center & FIC & National Agency & 1,248 & 25 \\
            Health Resources and Services Administration & HRSA & National Agency & 133 & 8 \\
            Indian Health Service & IHS & National Agency & 1,901 & 24 \\
            Michael J. Fox Foundation for Parkinson's Research & MJFFPR & Nonprofit Organization & 138 & 2 \\
            National Cancer Institute & NCI & National Agency & 39,009 & 2,830 \\
            National Eye Institute & NEI & National Agency & 1,031 & 9 \\
            National Heart, Lung, and Blood Institute & NHLBI & National Agency & 77 & 5 \\
            National Institute of Allergy and Infectious Diseases & NIAID & National Agency & 2,486 & 12 \\
            National Institute of Biomedical Imaging and Bioengineering & NIBIB & National Agency & 680 & 22 \\
            National Institute of Deafness and Other Communication Disorders & NIDOCD & National Agency & 122 & 44 \\
            National Institute of Dental and Craniofacial Research & NIDCR & National Agency & 235 & 3 \\
            National Institute of General Medical Sciences & NIGMS & National Agency & 1,076 & 57 \\
            National Institute of Health & NIH & National Agency & 16,618 & 718 \\
            National Institute of Mental Health & NIMH & National Agency & 1,126  & 82 \\
            National Institute on Minority Health and Health Disparities & NIOMHHD & National Agency & 211 & 3 \\
            Office of Inspector General & OIG & National Agency & 229 & 56 \\
            Pan American Health Organization & PAHO & International Organization & 5,967 & 3 \\
            Providence St. Joseph Health & PSJH & Nonprofit Organization & 77 & 3 \\
            St. Jude Children's Research Hospital & SJCRH & Nonprofit Organization & 9,247 & 22 \\
            Substance Abuse and Mental Health Services Administration & SAMHSA & National Agency & 180 & 143 \\
            World Food Programme & WFP & International Organization & 3,474 & 18 \\
      	\bottomrule
      \end{tabu}%
    \end{table*}

\section{Related Work}

This paper intersects research on real-world visualization corpora, empirical studies of visualization design, and its use in public health and high-stakes communication. We review these areas and highlight key studies that position our work within the expanding research landscape. For a broader overview of corpus-based research in visualization, we recommend the survey by Chen et al.~\cite{chen2023state}.


\subsection{Empirical Studies of Real-World Corpora}

Prior work in visualization has utilized large collections of real-world visualizations to study design, communication, and automated analysis (e.g., ~\cite{poco2017reverse,lee2017viziometrics,jung2017chartsense,chen2020composition,chen2021vis30k}). Borkin et al.’s study of over 2,000 visualizations identified key visual features, such as color, density, and imagery, that enhance memorability~\cite{borkin2013makes}. This work had a lasting impact through the release of the \textit{MassVis} dataset, which has become essential across various research areas. It facilitated large-scale studies of visual attention and eye fixation behavior in information visualizations~\cite{bylinskii2015eye}, and more recent research on communication, recall, and trust in visualization~\cite{arunkumar2025modeling, pandey2023you, mckinley2025trustworthy}. 

Subsequent work has expanded this research by creating large-scale visualization datasets and repositories, such as \textit{VizNet}, which offers a vast collection of visualizations and datasets for machine learning and analysis of visualization structures~\cite{hu2019viznet}. \textit{VisImages} enhanced this direction with a detailed, expert-annotated corpus for in-depth structural and semantic analysis of visualizations. \cite{deng2022visimages}. \textit{Beagle} showcased automated extraction and interpretation of web visualizations, facilitating large-scale recovery of visual artifacts beyond curated datasets \cite{battle2018beagle}. 

A parallel line of work has used real-world visualization collections to study public communication and misinformation. For instance, Zhang et al. mapped COVID-19 crisis visualizations to analyze their use during the public health emergency~\cite{zhang2021mapping}. Lee et al. explored how visualization practices in online discourse can support unconventional scientific claims by utilizing traditional conventions~\cite{lee2021viral}. Others examined ``misleading'' visualization practices to study how deception might manifest in real-world artifacts and communicative contexts \cite{lisnic2023misleading}. These studies show that real-world corpora can reveal structural properties of visualization design and their social and communicative roles.


\subsection{Embellishment} 

Visualization research has explored how design features beyond core data encoding affect interpretation and engagement. A key focus is on \textit{visual embellishments}, i.e.,  design elements unnecessary for decoding data and often linked to Tufte’s concept of "chartjunk" \cite{tufte1983visual}. These include color, direct labeling, iconography, imagery, and explanatory text.

Early perspectives focused on minimalist design with high data-ink ratios to enhance analytical clarity \cite{tufte1983visual}. However, research has shown that embellishment can also play a significant role. For example, Bateman et al. found that embellished charts improved recall without significantly impacting comprehension accuracy \cite{bateman2010useful}. Borkin et al. showed that color, recognizable objects, and high visual density contribute to memorability \cite{borkin2013makes}. Other works have shown that infographic-style visualizations are often perceived as more engaging and aesthetically appealing than plain charts \cite{andry2021interpreting}, and that design choices impact perceived credibility \cite{song2025visualizing} and clarity~\cite{pandey2023you}.

At the same time, prior work highlights important tradeoffs. Certain embellishments can reduce clarity, introduce ambiguity, or negatively affect trust depending on the visual style and communicative context \cite{andry2021interpreting, song2025visualizing}. For example, Song et al. found that comic-style visualizations and hand-drawn font treatments reduced credibility relative to more conventional visual forms \cite{song2025visualizing}. Similarly, Skau et al. found that shape-based embellishments did not improve performance over standard bar charts \cite{skau2015evaluation}. Borgo et al. found that visual embellishments improved information retention but increased processing times \cite{borgo_empirical_2012}. Thus, the effects of embellishment are highly context-dependent and often reflect a tension between engagement, memorability, and interpretability.

\subsection{Structure \&  Accessibility} 

Previous work highlights the need for structural clarity and accessibility in visualizations (e.g., ~\cite{joyner2022visualization,kim2021accessible,elavsky2022accessible}). Essential design elements such as axes, legends, labels, annotations, and color accessibility are vital for accurate interpretation and inclusive communication, particularly in public-facing, high-stakes contexts where simplicity, readability, and precision must be balanced~\cite{ottley2026consensus}.
Moreover, prior work has shown that titles, labels, annotations, and other structural elements influence viewers' attention, perceived informativeness, confidence, and interpretation of visualizations \cite{borkin_beyond_2016, arunkumar_image_2024, fung_effectiveness_2024}.

Accessibility has likewise become central to visualization research. Researchers have developed auditing frameworks such as \textit{Chartability} \cite{elavsky2022accessible}, evaluated accessibility tools and interaction techniques for people with low vision and blindness \cite{kim_beyond_2023}, and expanded the discussion to broader disability communities \cite{wimer_beyond_2024}. Accessibility is crucial in public health, as organizations must adhere to regulations like the Americans with Disabilities Act with the Web Content Accessibility Guidelines (WCAG). Our analysis centers on color vision deficiency (CVD) while supporting future efforts to expand accessibility evaluations.

\subsection{Visualization in Public Health Practice}

Research on data visualization in public health has mainly focused on applications and systems. Several studies have developed tools for tasks like disease surveillance, intervention targeting, cohort analysis, and public health communication~\cite{valdiserri2018data, sullivan2020data, cocoros2021riskscape, shaban2017pophr, ledesma2016health}. For example, Sullivan et al. introduced \textit{AIDSVu}, an interactive map for HIV prevention and care in the U.S. \cite{valdiserri2018data, sullivan2020data}. Cocoros et al. developed a visualization platform for public health surveillance in Massachusetts \cite{cocoros2021riskscape}. Other research has focused on software libraries for creating health-related visualizations \cite{sharma2023protocol, ledesma2016health}.

Research on visualization in public health includes literature surveys, scoping reviews, and practical analyses, focusing on decision-making, chart types, dashboard design, electronic health records, and health communication (e.g., \cite{park2022impact, ofori2025visual, narayan2021need, tan2025data, west2015innovative}). Ofori et al. surveyed 28 articles on high-level design choices for visualization tools in public health \cite{ofori2025visual}. Arleo et al. evaluated COVID-19 dashboard designs and provided expert recommendations \cite{arleo_reflections_2025}. Riddell et al. reviewed 84 papers on network methods and visualizations in public health data \cite{riddell_methods_2026}. Preim et al. examined visual analytic solutions in public health, highlighting techniques and use cases \cite{preim_survey_2020}. 

Most relevant to this work, Park et al. conducted a qualitative study with 14 public health officials to explore the methods, needs, and challenges of creating visualizations in practice \cite{park_understanding_2021}. Their findings highlighted the use of visualizations for communication with health department staff, policymakers, stakeholders, and the public, with a preference for simple, easily interpretable formats. Participants emphasized the effectiveness of infographics for narrative communication and behavior change, as well as maps for displaying geographic distributions of public health issues. However, they also faced challenges such as difficulty accessing appropriate data, inconsistent data sources, and limited staff expertise in visualization design.


\medskip
\noindent
In summary, prior research has explored visualization design through experiments and practitioner perspectives, but it \textit{lacks a systematic, large-scale understanding of actual design practices.} There's limited knowledge about how design features co-occur, how choices differ across organizations, and how real-world constraints affect visualization production. This work aims to fill that gap with an empirical analysis of public health visualizations.

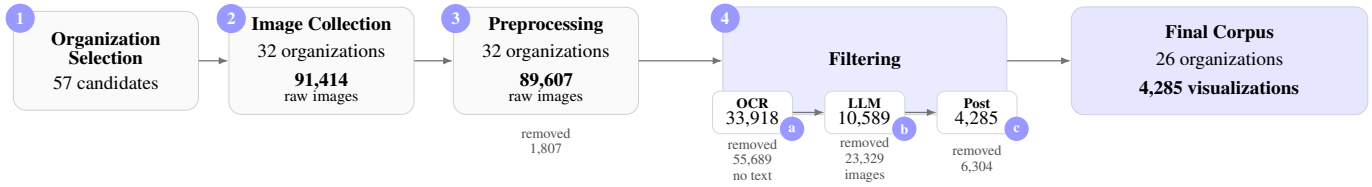
\begin{figure*}[t]
\centering
\begin{adjustbox}{width=\linewidth,center}
\begin{tikzpicture}[
  major/.style={
    draw=black!18,
    rounded corners=8pt,
    fill=black!2,
    minimum width=4.0cm,
    minimum height=2.2cm,
    text width=3.7cm,
    align=center,
    inner sep=6pt
  },
  majorblue/.style={
    draw=blue!35!black!18,
    rounded corners=8pt,
    fill=blue!8,
    minimum width=6.3cm,
    minimum height=2.4cm,
    text width=5.9cm,
    align=center,
    inner sep=6pt
  },
  majorfinal/.style={
    draw=blue!35!black!18,
    rounded corners=8pt,
    fill=blue!10,
    minimum width=6.6cm,
    minimum height=2.4cm,
    text width=6.1cm,
    align=center,
    inner sep=6pt
  },
  subbox/.style={
    draw=black!16,
    rounded corners=5pt,
    fill=white,
    minimum width=1.75cm,
    minimum height=0.95cm,
    text width=1.5cm,
    align=center,
    inner sep=4pt
  },
  arrow/.style={
    -{Latex[length=2.2mm]},
    thick,
    draw=black!55
  },
  note/.style={
    font=\normalsize,
    text=black!68,
    align=center
  },
  count/.style={
    font=\footnotesize\bfseries,
    text=black!72
  }
]

\node[major] (sel) at (0,0)
{\textbf{\Large Organization Selection}\\[6pt]
{\Large 57 candidates}\\[6pt]};

\node[major] (collect) at (4.8,0)
{\textbf{\Large Image Collection}\\[6pt]
{\Large 32 organizations}\\[6pt]
{\bfseries\Large 91,414}\\
\large raw images};

\node[major] (prep) at (9.8,0)
{\textbf{\Large Preprocessing}\\[6pt]
{\Large 32 organizations}\\[6pt]
{\bfseries\Large 89,607}\\
\large raw images};

\node[majorblue] (filter) at (16.9,0)
{\textbf{\Large Filtering}\\[10pt]};

\node[majorfinal] (final) at (24.8,0)
{\textbf{\Large Final Corpus}\\[6pt]
{\Large 26 organizations}\\[6pt]
{\bfseries\Large 4,285 visualizations}};
\draw[arrow] (sel) -- (collect);
\draw[arrow] (collect) -- (prep);
\draw[arrow] (prep) -- (filter);
\draw[arrow] (filter) -- (final);

\node[note] at (9.8,-1.80){removed\\1,807};

\node[subbox] (ocr) at (14.4,-1.15)
{\textbf{OCR}\\{\Large 33,918}};

\node[subbox] (llm) at (16.9,-1.15)
{\textbf{LLM}\\{\Large 10,589}};

\node[subbox] (post) at (19.4,-1.15)
{\textbf{Post}\\{\Large 4,285}};

\draw[arrow] (ocr) -- (llm);
\draw[arrow] (llm) -- (post);

\node[note] at (14.4,-2.20) {removed\\55,689 \\no text};
\node[note] at (16.9,-2.20) {removed\\23,329\\images};
\node[note] at (19.4,-2.20) {removed\\6,304};


\node[anchor=north west, scale=1.2] at ($(sel.north west)+(-0.3,0.3)$) {\circnum{1}};
\node[anchor=north west, scale=1.2] at ($(collect.north west)+(-0.4,0.2)$) {\circnum{2}};
\node[anchor=north west, scale=1.2] at ($(prep.north west)+(-0.4,0.2)$) {\circnum{3}};
\node[anchor=north west, scale=1.2] at ($(filter.north west)+(-0.4,0.2)$) {\circnum{4}};
\node[anchor=north west] at ($(ocr.south east)+(-0.4,0.5)$) {\circnum{a}};
\node[anchor=north west] at ($(llm.south east)+(-0.4,0.5)$) {\circnum{b}};
\node[anchor=north west] at ($(post.south east)+(-0.4,0.5)$) {\circnum{c}};

\end{tikzpicture}
\end{adjustbox}
\vspace{-.3cm}
\caption{Corpus construction and annotation pipeline. The top-level stages align with the methodological subsections, while nested boxes preserve step-level detail. Starting from 57 candidate public health organizations, we scraped 91,414 raw images and progressively reduced noise through preprocessing and multi-stage filtering, yielding a final corpus of 4,285 visualizations from 26 organizations.}
\label{fig:corpus_pipeline}
\vspace{-.3cm}
\end{figure*}

\section{Method}
We collected a large-scale, ecologically grounded corpus and empirically characterized dominant chart forms, communication tasks, structural conventions, and accessibility issues. We gathered \textbf{91,414} candidate images and refined them to \textbf{4,285 visualizations} through a semi-automated filtering process. We developed a codebook, annotated the corpus using GPT-5 and complementary algorithms, and manually validated random samples after each major processing stage. Figure ~\ref{fig:corpus_pipeline} summarizes the collection and filtering pipeline.


\subsection{Corpus Scope and Source Selection}
\stepX{1}We began by compiling an initial candidate set of \textbf{57 public health organizations} representing a diverse cross-section of public-facing health communication, including nonprofit organizations, international and multilateral agencies, and U.S. national public health institutions. We selected U.S. and international organizations because of their global influence and their relevance to the literature reviewed. To construct this sampling frame, we conducted targeted Google searches and drew organizations from publicly available directories. We used five directories, including the 2025 Forbes Top Charities and NIH Institutes and Centers; see Appendix~\ref{appendix:orgs_sources} for the complete list of sources and URLs. Organizations were included only if their primary mission focused on health, medicine, or public health.


The resulting candidate set included \Nnonprofit nonprofit organizations, \Ninternational international and multilateral organizations, and \NUS U.S. national agencies. For the purposes of this study, we treated sub-agencies as distinct organizations when they maintained separate public-facing websites and independently curated visual communication materials. For example, the National Institutes of Health (NIH) and the National Institute of Dental and Craniofacial Research (NIDCR) were treated as separate organizations because they operate distinct websites and publish separate visual content, even though NIH is the parent agency of NIDCR. We then applied exclusion criteria to retain only organizations that could be systematically incorporated into the corpus. Organizations were excluded if their websites were inaccessible due to security restrictions or firewalls ($n=14$), incompatible with our scraping pipeline (e.g., dynamic content, unsupported rendering frameworks, or non-standard image delivery mechanisms; $n=5$), or yielded no images after automated collection ($n=6$). After applying these criteria, the set consisted of \textbf{32} organizations that were successfully incorporated into the image corpus.

\subsection{Web Collection and Image Retrieval Pipeline}
\stepX{2}To support consistent retrieval across heterogeneous website structures, we used a targeted keyword-based search strategy established on terms commonly associated with data communication: \textit{data}, \textit{chart}, \textit{map}, \textit{visualization}, \textit{graph}, \textit{infographic}, \textit{dashboard}, and \textit{plot}. Each keyword was entered into the search function of every organization's website, and all resulting directory index pages were manually collected. For example, searching \textit{data} on the Alzheimer's Association website required manually collecting subsequent index pages (e.g., pages 2, 3, etc.). This process yielded a total of \textbf{10,246 candidate URLs}. The scraper then operated iteratively in three stages. First, it visited each collected index page URL. Second, it recursively opened all linked listing pages and content pages accessible from those directories. Third, it extracted all embedded image assets from each visited page.

Data collection was conducted \textbf{from January to February 2026} using our self-designed web scraper. Across this process, we retrieved a total of \textbf{91,414 candidate images}. In addition to image assets, we collected associated webpage metadata when available, including the page title, description, source URL, page author, and last modified date.
These metadata were retained to support future analyses, and summarized in Appendix~\ref{appendix:meta_data}.

\subsection{File Preparation and Corruption Checks}
\stepX{3}We used the Python Imaging Library (PIL) and \texttt{svglib} to standardize image formats and remove unusable files to support downstream computational processing.
SVG and WEBP file formats were converted to PNG to ensure compatibility across subsequent analysis steps. We additionally identified and removed corrupted image files that could not be successfully opened or processed. After this stage, 2,522 files were converted and \textbf{1,807 images} were removed due to corruption or format issues, yielding \textbf{89,607 images} for subsequent analysis.

\subsection{Visualization Identification and Multi-Stage Filtering}
\stepX{4} Because public-facing organizational websites contain a substantial number of non-visualization images, the raw image collection included significant noise. The majority of retrieved images were not data visualizations, which is consistent with prior work on web scraping public-sector and news websites~\cite{borkin2013makes, lo_misinformed_2022}. The collected noise in our study was software screenshots, logos, promotional flyers, anatomical diagrams, document pages, organizational charts, and photographs. To construct a high-quality visualization corpus, we implemented a four-stage filtering pipeline that progressively reduces noise while preserving candidate visualizations.

\stepX{4a}\textbf{OCR-Based Text Presence Filtering.} We applied an optical character recognition (OCR) filter, assuming that data visualizations contain some textual elements, such as axis labels, legends, titles, tick labels, or direct annotations. This stage was designed primarily to remove photographs and other non-informational imagery, which we anticipated to be a major source of the collected noise. We used \texttt{Pytesseract}, a Python wrapper for Google’s Tesseract OCR engine, to detect text within each image. Images for which no text was detected were removed from the candidate corpus. Additional removals occurred 1) if the image structure wasn't decipherable by OCR, or 2) if the OCR text extraction was empty after applying text rules to remove special characters and spaces. This step removed \textbf{55,689 images}, reducing the corpus to \textbf{33,918 candidate images}. All extracted OCR text was retained for later use in taxonomy development and metadata enrichment.

\smallskip\noindent
\textit{Validation}\hspace{0.1em} An author manually reviewed a random sample of \textbf{406} images, achieving \textbf{84\%} agreement with the automated decision. Disagreements typically occurred because images exhibited bad resolution, poor contrast, or were too small, which would've removed in other stages of the pipeline.


\stepX{4b}\textbf{LLM-Assisted Visualization Identification.} We used OpenAI’s \texttt{gpt-5-mini} model to classify each candidate image using a structured prompt that included the following outputs: whether the image was a visualization, chart type labels, text presence, noise category, structural features (e.g., axes, marks, panels), and a confidence score ranging from 0 to 1. To maximize precision, we retained only images that satisfied all of the following criteria: (1) labeled as a visualization, (2) assigned at least one valid chart type, (3) noise category labeled as \texttt{none} or \texttt{other}, and (4) model confidence greater than \textbf{0.85}. We chose a strict threshold of 85\% to have greater confidence that noise was removed from our corpus, but we note that the trade-off is the potential exclusion of valid visualizations.
This step removed \textbf{23,329 images}, yielding \textbf{10,589 candidate visualizations}.

\smallskip\noindent
\textit{Validation}\hspace{0.1em} An author manually validated a random sample of \textbf{100 images}, with agreement the image contains a data visualization exceeding \textbf{97\%}. Labels generated during this stage were retained for later design-space analysis. Disagreement typically occurred with hierarchical organizational diagrams, which were intended to be classified as noise but were assigned as a visualization from the LLM. 


\stepX{4c}\textbf{Post-Processing and Deduplication.} Finally, we applied a post-processing stage to remove residual artifacts that persisted through earlier filters. Images were excluded if they were classified as (1) too small or (2) a duplicate. 1) We considered an image too small if both width and height were less than or equal to \textbf{400 pixels}, which, depending on the device, is comparable to the size of a postage stamp. 2) We used \texttt{PIL} and the \texttt{imagehash} packages to identify exact duplicates using perceptual hashing and retained a single canonical copy for each unique image. 24 images were manually removed because the related organizations met the exclusion criteria described in Section 3.1; the image artifacts were collected before the organizations were observed to meet the exclusion criteria. This final stage removed \textbf{6,304 images}, resulting in a final corpus of \textbf{4,285 visualizations}. Another \textbf{6 organizations} were ultimately removed because their image collections became empty after the reduction process. The final corpus, therefore, consists of \textbf{26 public health organizations} and \textbf{4,285 visualizations}. Table~\ref{tab:org_counts} summarizes the organization-level counts from initial retrieval through final corpus inclusion.






\subsection{Codebook Development}

We used an iterative codebook development process.
Instead of directly applying labels from prior work in visualization taxonomy, the authors conducted an open-coding pass to ensure appropriate capture of design elements in this specialized collection of visualizations. One author conducted an initial open-coding pass over the sampled visualizations, documenting all meaningful annotation attributes observed in each artifact. The sample was constructed by randomly selecting 5 visualizations (if applicable) from each organization in the corpus, resulting in a total of 131 visualizations reviewed. These attributes spanned multiple dimensions of visualization design, including content (e.g., demographic focus such as gender or race), visual embellishments (e.g., iconography, photographs, decorative elements), and structural features (e.g., chart type, axis labels, legends, annotations).  This process initially yielded 30 variables

The full author team then reviewed and iteratively refined the initial codes into a consolidated codebook. During this process, redundant, overly specific, or semantically overlapping attributes were merged into broader coding dimensions. For example, rotated labels, font size, and color scale tags were removed because they were overly specific or unlikely to be consistently captured across the corpus. Rather than annotating individual data types (e.g., categorical, time series, quantitative), we developed a broader \textit{Data Context} taxonomy to characterize the visualization's primary communicative task.

The final annotation schema comprised 22 variables organized into six categories: \textit{chart family}, \textit{data context}, \textit{structural features}, \textit{visual embellishments}, \textit{color}, and \textit{text}. Table~\ref{tab:codebook} summarizes the codebook and details the variables in the Appendix ~\ref{appendix:codebook}.
The resulting codebook was used to annotate the full corpus and support the subsequent design-space analyses.

\subsection{LLM-Assisted Corpus Annotation and Validation}

Following codebook development, we operationalized the final annotation schema as two structured prompting frameworks and applied it to the full corpus using OpenAI’s \texttt{gpt-5-mini} model. The prompts used a controlled vocabulary derived directly from the finalized codebook. The first prompt classified each visualization with \textit{visual embellishments} and \textit{color} features. \textit{Visual embellishments} features used a binary feature schema (e.g., presence of attributes such as iconography). The \textit{color} associated attribute within this prompt was the extraction of the HEX values from the visualization. The second prompt classified each visualization according to \textit{chart family}, \textit{data context}, \textit{structural features}, and a confidence score ranging from 0 to 1. To account for hybrid visual forms, the model was instructed to return both a \textit{primary} and an optional \textit{secondary} chart type. Likewise, the model selected \textit{primary} and \textit{secondary} options for \textit{data context}. To ensure consistency, \textit{structural features} and \textit{data context} attributes were accompanied by explicit natural-language definitions in the prompt. Like visual embellishments, \textit{structural features} attributes follow a binary feature schema (e.g., denoting the presence of direct labels, legends, and grid lines). 

The model was required to return all outputs as structured JSON. Some fields were excluded from subsequent analyses. Secondary chart-family labels were omitted because GPT assigned over 50\% of visualizations to the generic ``Other'' category, preventing meaningful analysis of hybrid visual forms. For consistency, secondary data-context labels were also excluded. The confidence score from the second annotation prompt was removed from forthcoming analyses. We adopted a conservative confidence threshold of 87\% to guide quality assurance. We emphasize that this threshold was used only to prioritize manual review and not as a decision boundary for the final dataset. Because over 75\% of visualizations exceeded this threshold, confidence scores were not included in the following analyses.
The final annotated corpus contains structured labels for chart family, data context, structural features, visual embellishments, and color values for all \textbf{4,285} visualizations. \emph{The corpus, codebook, GPT prompts, analysis scripts, and archived URLs are available} at \url{https://washuvis.github.io/ChartingPublicHealth/}.


\medskip\noindent
\textit{Validation}\hspace{1em} 
An author manually reviewed a random sample of \textbf{100} visualizations in the final corpus. Agreement was assessed for features relating to \textit{chart family}, \textit{data context}, \textit{structural features}, \textit{visual embellishments}, and \textit{color}. Across the validation sample, agreement was \textbf{98\%} for primary chart type, \textbf{98\%} for primary data context, and \textbf{98\%} for HEX colors. In consideration of all structural features and visual embellishments attributes, \textbf{95\%} average agreement for structural features, and \textbf{90.6\%} average agreement for visual embellishments. Regarding visual embellishments, disagreements typically arose from the over-application of a particular feature from the GPT model. For instance, visualizations that listed the data sources were sometimes denoted as having text annotations. 


\subsection{Derived Measures and Communication Risk Indicator}
We derived a set of quantitative measures designed to capture higher-level properties of visualization communication and accessibility, such as the concentration of embedded text and potential accessibility risks associated with color palette selection. 

\subsubsection{Text Density.}
\label{sec:text_density}
To quantify the amount of written information embedded in each visualization, we computed a normalized text density measure based on optical character recognition (OCR). For each image, text was first extracted using Tesseract OCR (as described in Section 3.4). To reduce noise in our text density computation, symbols, whitespaces, punctuation, and numbers were removed from the extracted text. We created tokens by aggregating all the characters left in the text. We then normalized text quantity by image area (width and height) to account for differences in visualization size:

\[
D_{\text{text}} = \frac{N_{\text{char}}}{W \times H}
\]
\noindent
where $N_{\text{char}}$ is the total number of retained text characters, and $W$ and $H$ denote the image width and height in pixels, respectively.
This normalization yields a unitless measure of text density that enables comparison across visualizations of varying dimensions, where larger values indicate a greater concentration of embedded textual information relative to image size.

\subsubsection{Color Blindness Risk}
\label{sec:color_risk}
To assess potential accessibility concerns, we derived a binary \textit{color blindness risk} indicator for each visualization based on the extracted color palette. For each unique set of detected hex color values, we evaluated whether the palette remained perceptually distinguishable under normal vision and common forms of color vision deficiency (CVD), including protanomaly, deuteranomaly, and tritanomaly.

Each palette was first converted from hexadecimal color codes into normalized sRGB values. We then simulated three common CVD conditions using a perceptual color space transformation implemented with \texttt{colorspacious}. For each condition, we computed pairwise Euclidean distances between all color pairs in CIELAB space, using $\Delta E_{00}$, which provides a perceptually grounded estimate of color distinguishability by accounting for differences in lightness, chroma, and hue.

For each visualization, we determined palette separability using the minimum pairwise perceptual distance between all colors:

\[
d_{\min} = \min_{i,j,c} \Delta E_{00}^{(c)}(i,j)
\]

\noindent
where $i$ and $j$ index color pairs and $c$ indexes normal vision and the three simulated CVD conditions.

A palette was classified as \textit{risky} if at least one pair of colors exhibited limited perceptual separation under any viewing condition. Following common practical interpretation thresholds, we considered palettes with $\Delta E_{00} < 5$ to indicate elevated accessibility risk, and those with $\Delta E_{00} < 2$ to indicate near-indistinguishable color pairs. See Figure \ref{fig:color_accessibility} in Appendix \ref{appendix:color_blind} for more details.

For the binary indicator used in our primary analysis, a visualization was marked as risky if:
\[
R =
\begin{cases}
1 & \text{if } d_{\min} < 5 \\
0 & \text{otherwise}
\end{cases}
\]

\medskip\noindent




\begin{figure*}[!t]
    \centering
    \includegraphics[width=\linewidth,height=0.7\textheight, alt={Figure 3 consists of multiple bar charts that describe the percentage rate of each feature in the entire 4,285 visualization collection. Important items are discussed in Section 4.}, keepaspectratio]{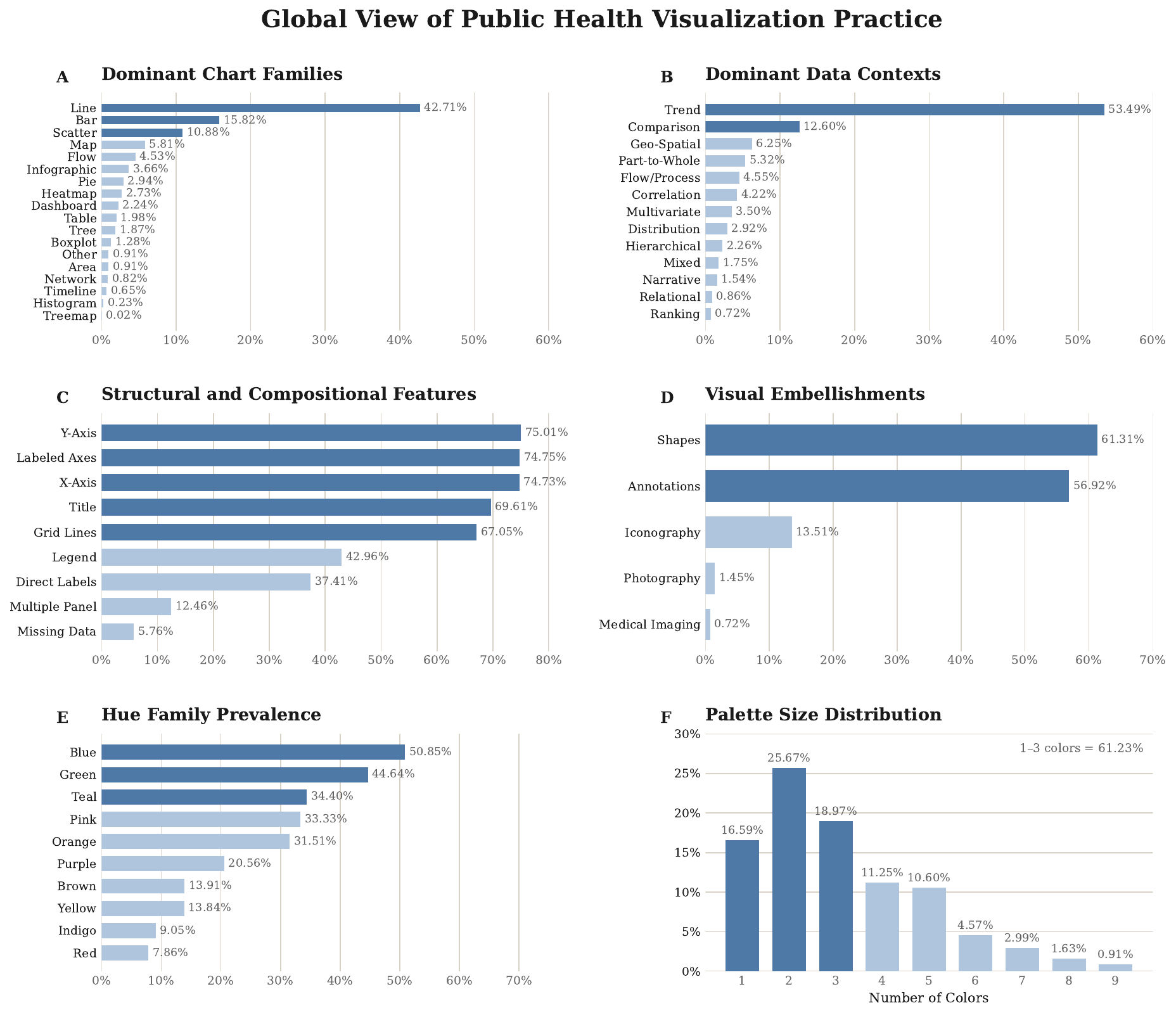}
    \vspace{-.3cm}
    \caption{A summary of the analysis findings. Line charts and temporal trends are most prevalent in the corpus, suggesting a strong emphasis on communicating change over time. Chart elements such as axes, labels, and titles are present in most visualizations, while direct labeling is less common. Text annotations and shape/line-based embellishments appear in most charts, indicating that ``embellishments'' are common in practice.}
    \vspace{-.3cm}
    \label{fig:global_overview}
\end{figure*}

\section{Corpus-Level Results}
In this section, we characterize the visualization design space through analyses of chart families, data contexts, structural features, visual styling, and derived communication measures, including text density, color usage, and accessibility risk.

\subsection{A Small Set of Chart Types}
As shown in Figure~\ref{fig:global_overview}A, line charts account for 42.71\% of the corpus, substantially exceeding all other chart types. Bar charts (15.82\%) and scatter plots (10.88\%) are the next most prevalent forms. Together, these three chart families comprise nearly 70\% of the entire corpus.
The remaining design space is comparatively sparse. Maps account for 5.81\% of visualizations. Infographics (3.66\%), dashboards (2.24\%), tables (1.98\%), and more structurally specialized representations such as treemaps (0.02\%) appear infrequently. This long-tailed distribution suggests that public health communication relies heavily on a small, stable set of familiar chart forms.
\textit{We observe limited chart diversity, with organizations overwhelmingly favoring conventional charts over more specialized or exploratory designs.}

\subsection{Temporal Trend Communication Dominates}
We can observe from Figure~\ref{fig:global_overview}B that the dominant chart families align closely with the primary data contexts observed in the corpus. Temporal trend communication is by far the most common analytical context, accounting for 53.49\% of all visualizations. This prevalence substantially exceeds the categorical comparison (12.6\%) and the geographic context (6.25\%).
Other contexts, including part-to-whole composition (5.32\%), flow processes (4.55\%), and correlation relationships (4.22\%), occur substantially less frequently. More complex contexts, such as hierarchical (2.26\%) and mixed multi-context visualizations (1.75\%), are rare.
\textit{The corpus is strongly oriented toward communicating change over time, consistent with the prevalence of line charts.}

\subsection{Obscurity in Structural Conventions}
Structural elements associated with readability and interpretability appear frequently but not universally. Figure~\ref{fig:global_overview}C shows that titles are present in 69.61\% of visualizations, and grid lines appear in 67.05\%. Axes and axis labels are present in approximately three-quarters of the corpus, with Y-axis presence at 75.01\%, X-axis presence at 74.73\%, and labeled axes at 74.75\%.
However, other critical interpretive structures appear considerably less often. Legends are present in only 42.96\% of visualizations, and direct labels in 37.41\%. In visualization design, they often occur in opposition of each other, e.g., legend or direct labeling, and to have both present may be seen as redundant. This is demonstrated by 69\% of visualizations having direct labeling or legends, and approximately 12\% exhibiting both features. Multiple-panel designs remain relatively uncommon (12.46\%), and explicit denotation of missing data appears in only 5.76\% of cases.
\textit{While public health organizations generally adhere to core structural conventions, the limited use of legends and direct labels may create opportunities for interpretive ambiguity in public-facing communication.}

\begin{figure*}[!t]
    \centering
    \includegraphics[width=.95\linewidth, alt={Figure 4. features a heatmap and scatterplot on the PCA clusters and their features. Key details are described in Section 5.}]{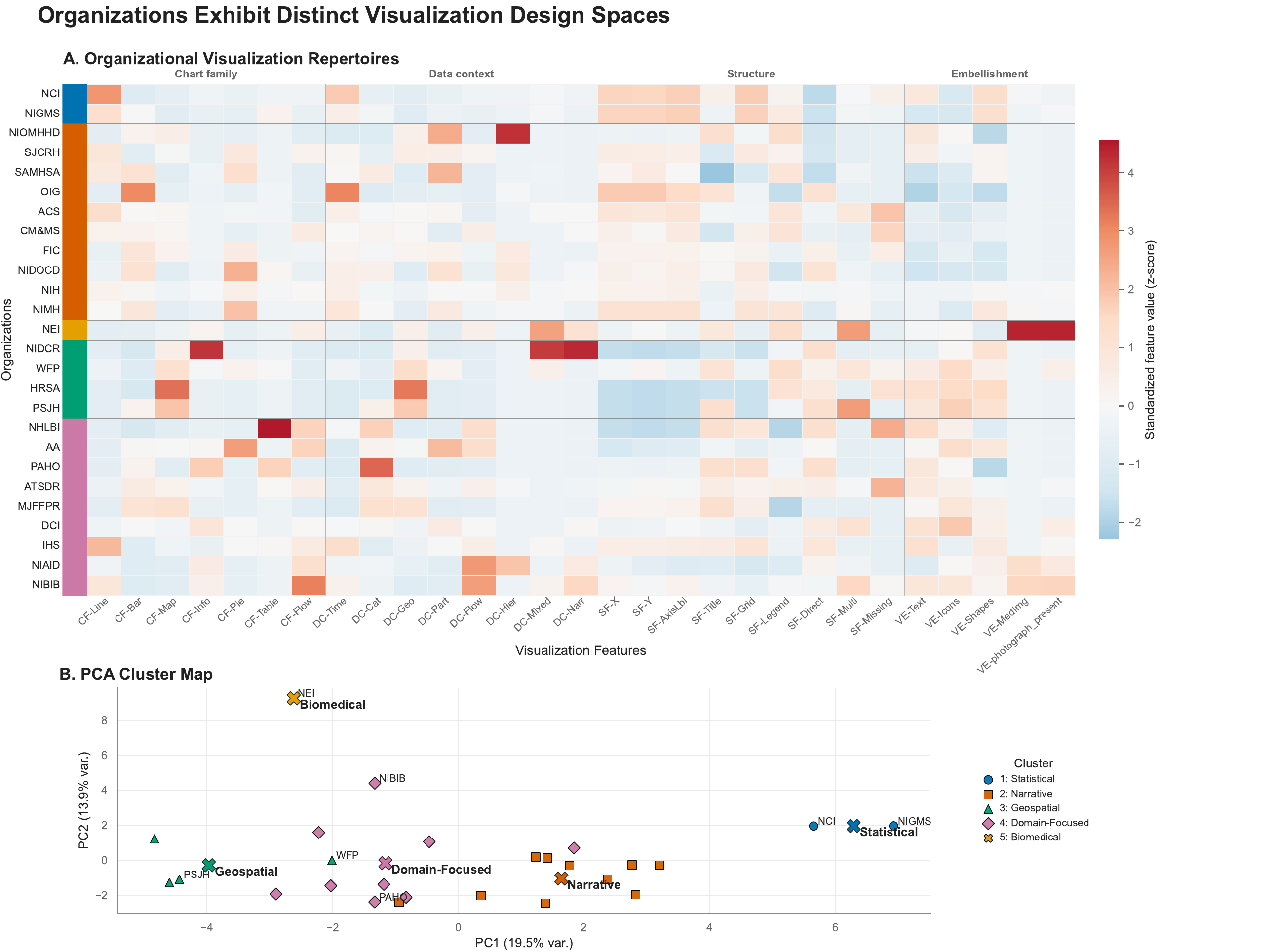}
    \caption{
    Organizational visualization design space revealed through hierarchical clustering.
    (A) Clustered heatmap of standardized feature prevalence across chart family (CF), data context (DC), structural features (SF), visual embellishments (VE), and text density (TD).
    (B) PCA projection of the same feature space, showing separation among the five identified clusters.
    }
    \label{fig:org_clusters}
    \vspace{-.4cm}
\end{figure*}

\subsection{Embellishments Are Functional}
Figure~\ref{fig:global_overview}D shows that lightweight embellishments are common, particularly unique point shapes and line markings (61.31\%) and text annotations (56.92\%), suggesting that organizations frequently use minimal stylistic variation to support differentiation and emphasis. More expressive embellishments, however, remain uncommon. For example, iconography appears in only 13.51\% of visualizations. 

Using the normalized text density measure defined in Section~\ref{sec:text_density}, we examined text usage in the corpus. Text density varied substantially across the corpus, suggesting notable differences in communication style across organizations. The bottom 25\% of visualizations contained very little embedded text ($\leq 1.33\times10^{-4}$), indicating highly minimalist designs. By contrast, only the top 25\% of visualizations exceeded a text density of $1.70\times10^{-3}$, suggesting that text-rich explanatory designs were comparatively uncommon. Most visualizations balanced graphical encoding with modest amounts of supporting text,
\textit{exhibiting a restrained but consistent use of visual styling
features.}

\subsection{Color Usage Reflects Strong Palette Conventions}
Figure~\ref{fig:global_overview}E suggests that color usage across the corpus reveals strong and highly consistent palette conventions. Blue is by far the most prevalent hue, appearing in 50.85\% of all visualizations, followed by green (44.64\%) and teal (34.4\%). Secondary colors such as pink (33.33\%), orange (31.51\%), and purple (20.56\%) also appear frequently, while warm and high-salience hues such as red (7.86\%) and yellow (13.84\%) are comparatively rare. In addition to hue selection, palette size is similarly constrained. Our computed palette size only considers hue color, e.g., blue, green, or teal, and not grayscale-associated colors, e.g., gray, white, or black.  We can observe from Figure~\ref{fig:global_overview}F that the most common palette sizes are two-color (25.67\%) and three-color (18.97\%) designs, followed by single-color visualizations (16.59\%). Together, visualizations using one to three colors account for over 61\% of the corpus. Larger palettes are substantially less common, with palettes of more than 5 colors occurring infrequently. 
In sum, the visualizations in the corpus rely heavily on cool-toned palettes, particularly blue and green hues, which may reflect both institutional branding conventions or the perceived neutrality, professionalism, and trustworthiness often associated with these colors\cite{su2019trustworthy,elliot2015color,mehta2009blue}. \textit{Consistent with the previous analyses, the limited chromatic complexity indicates a preference for simplicity in public health visualization design.}

\subsection{Potential Color-Blind Accessibility Risks}
Because public health visualizations are often designed for broad public audiences, accessibility is a critical dimension of communication effectiveness. In particular, inadequate color separation can substantially reduce interpretability for viewers with common forms of color vision deficiency, potentially obscuring trends, categories, or comparisons that are central to decision-making. To assess how frequently such risks appear in practice, we evaluated the corpus using the \textit{color blindness risk} measure described in Section~\ref{sec:color_risk}. 


We found that color blind accessibility concerns persist in a nontrivial portion of the corpus. Using a practical perceptual distinguishability threshold of $\Delta E_{00} < 5$, \textit{23.08\%} of visualizations contained at least one potentially ambiguous color pair under normal vision or simulated color vision deficiency conditions. Even under a stricter threshold of $\Delta E_{00} < 2$, \textit{6.18\%} of visualizations contained near-indistinguishable color pairs.
\textit{Together, these findings suggest that color-based ambiguity remains a meaningful accessibility concern in real-world public health visualizations, particularly for viewers with color vision deficiencies.}

\section{Organization-Level Design Profiles}
To examine whether organizations exhibit distinct visualization practices, we clustered each organization's aggregated visualization profile. For each organization, we computed the proportion of visualizations exhibiting each \textit{chart family} (CF), \textit{data context} (DC), \textit{structural feature} (SF), and \textit{visual embellishment} (VE). \textit{Color} and \textit{text} variables were excluded because they represent derived measures rather than GPT-assigned annotations. This produced an organization-by-feature matrix describing each organization's visualization practices.

All features were standardized (z-scores) prior to clustering. We then applied hierarchical agglomerative clustering with Ward linkage to identify \emph{organizational visualization styles}. A five-cluster solution was selected based on dendrogram structure and interpretability, as lower cut points merged distinct styles while higher cut points primarily over-segmented existing ones.

\subsection{Organizations Exhibit Distinct Design Profiles}

Figure~\ref{fig:org_clusters} shows the resulting clustered heatmap and a PCA projection of the same standardized feature space. Together, these views reveal \textbf{five distinct organization-level design profiles}, suggesting that institutions exhibit stable and interpretable visualization styles. Cluster labels are \statistical, \narrative, \geospatial, \domain, and \biomedical. The labels are descriptive and based on a visual inspection of representative charts. The underlying clusters were formed using low-level visual features extracted from the figures, while the labels (e.g., Statistical and Narrative) summarize the dominant visual characteristics observed within each cluster. In contrast, the PCA projection provides a complementary view of similarity at the organizational level, which does not necessarily align perfectly with the visually derived clusters. Figure~\ref{fig:profile_examples} showcases example visualizations for each design profile in Appendix~\ref{appendix:vis_examples}.

\smallskip\noindent
\statistical \ The first cluster comprises of \textbf{2,887 visualizations} across \textbf{2 organizations}, NCI and NIGMS. Visualizations in this profile exhibit features related to statistical analysis, e.g., regression lines, confidence intervals, p-values, and statistical significance annotations. In consideration of chart family, \statistical\ has a strong preference for \textit{line charts} at 55.6\%, with scatter plots being the second dominant choice at 15.2\%. The salient data context is \textit{temporal trends} at 63.9\%, which is a rate similar to the preferred chart forms. The design profile values \textit{high structural completeness}, with the labeled axes (85.9\%), x-axis (85.2\%), y-axis (85.1\%), grid lines (78.7\%), and titles (77.1\%) all above 75\%. \textit{Pictorial-related visual embellishments are rare.} Instead, their preferred embellishment features are unique point shapes and marked lines at 71.3\% and text annotations at 65.4\%. The prevalence of the other visual embellishment features is below 9\%. \textit{Overall, this design profile emphasizes standardized statistical reporting and an evidence-based communication style.}

\smallskip\noindent 
\narrative \ The second cluster comprises \textbf{1,155 visualizations} and \textbf{10 organizations}, including NIH, SJCRH, and ACS. Visualizations in this profile demonstrate design decisions related to mass communication and narrative. This is shown in its \textit{wide range of chart families} (e.g., flow diagrams, tree diagrams, infographics, and dashboards) and the use of explanatory elements (e.g., labels, legends, and text annotations). Its most prevalent chart family is a \textit{bar chart} at 37.3\%, followed by line charts at 17.3\%. In consideration of data contexts, \textit{temporal trends} and categorical comparison are favored at 35.5\% and 21\%, respectively. A \textit{variety of structural features are exhibited}, with direct labeling (60\%), the presence of y and x axes ($\approx$57\%), axis labels (54.7\%), title (49.9\%), grid lines (44.6\%), and legend (39.3\%) approximately above or equal to 40\%. Like structural features, the \textit{use of visual embellishments is broad}, with unique point shapes and marked lines at 39.9\%, text annotations at 34.2\%, and iconography at 20.3\%. \textit{The design profile utilizes a broad range of chart forms, structural features, and visual embellishments with the purpose of using the most fitting combinations to support the high-level narrative from the data. This further suggests that the \narrative \ communications are produced for a myriad audience, e.g., stakeholders, staff, and the general public.}

\smallskip\noindent 
\geospatial \ The third cluster comprises \textbf{32 visualizations} and consists of \textbf{4 organizations}: WFP, PSJH, NIDCR, and HRSA. Its visualizations predominantly relate to geospatial data, resulting in various map types (e.g., regional, choropleth, weather tracking, and facility). As mentioned, \textit{maps} and \textit{geographic spatial} are the salient chart family and data context \textit{geographic spatial} features at 62.5\% and 65.6\%, respectively. Its predominant structural features are \textit{direct labeling} at 84.4\% and \textit{titles} at 71.9\%. The profile utilizes \textit{visual embellishments consistently}: text annotation prevalence is at 62.5\% with unique point shapes and marked lines and iconography both at 59.4\%. \textit{The design profile predominantly consists of map visualizations and utilizes direct labeling and visual embellishments, e.g., text annotations and iconography, for clarity in their communication practices.} 

\smallskip\noindent
\domain \ The cluster comprises \textbf{202 visualizations} and consists of \textbf{9 organizations}: NIBIB, IHS, and PAHO. Under visual inspection, we observed that visualizations in this profile represent domain-specific phenomena (e.g., MRI images, spectra, movement tracking, microscopy, and pathology), suggesting that the intended audience is limited to the organization and experts related to their specific health topic. The profile employs \textit{a broad range of chart forms}, with \textit{bar charts} (27.7\%) and \textit{line charts} (12.4\%) as most prevalent. Favored data contexts are \textit{categorical comparison and geographic spatial} at 31.2\% and 19.3\%, respectively. The profile exhibits a wide range of structural features, but has a strong preference for \textit{direct labeling} at 81.2\% and the \textit{presence of titles} at 74.3\%. Like structural features, a variety of visual embellishments are used but greatly favor \textit{text annotations} at 64.9\%. Like \narrative, the design profile utilizes a variety of chart forms, structural features, and embellishments, but implores explanatory design elements (e.g., text annotations and direct labeling) at an increased rate. \textit{The profile characteristics reflect scientific communication that balances standardized presentation with attention-guiding elements.}

\smallskip\noindent

Finally, \biomedical \ consists of \textbf{9 visualizations} and \textbf{1 organization}: the National Eye Institute (NEI), indicating a highly distinctive visualization profile. It is characterized by frequent use of \textit{heatmaps} (33.3\%) and the data context of \textit{geographic spatial} (33.3\%). Salient structural features are  \textit{titles (89\%)}, \textit{multi-panel layouts (67\%)}, and \textit{legends (67\%)}. The profile's favored visual embellishment is \textit{text annotations} (44.4\%). Its defining characteristic is the use of \textit{medical imaging (22\%)} distinguishing it from all other design profiles and reflecting its specialized communication needs.  


\smallskip\noindent
Importantly, these profiles are not determined solely by subject matter. For example, cancer-related organizations span multiple clusters, indicating that organizations develop distinct ways of visualizing that reflect communication norms, intended audiences, and reporting goals in addition to the underlying content.

\subsection{Diverse Color Practices}
\textit{Color} features were not included in our PCA groupings because they are secondary measures; however, we still evaluated their consistency and contrast across all organizations in the corpus. We evaluated the presence of each hue category (e.g., blue, green, and indigo) across all visualizations of an organization. 23\% of organizations use a particular color across all visualizations, indicating a consistent color palette or strong brand identity, such as blue for the Health Resources and Services Administration (HRSA) and orange for the Michael J. Fox Foundation for Parkinson's Research (MJFFPR). Blue is the dominant color choice for 73\% of organizations. The size of the color palette in a visualization typically varies across an organization's corpus, with only a few demonstrating strong consistency in palette size. The Office of Inspector General (OIG) exhibits the highest consistency in color palette size, with 91\% of visualizations using a single hue. Grayscale visualizations are rarely produced across organizations, and many don't have a single one. \textit{Organizations' consistency in color selection shows that color may be used to enforce a visual identity.}

    
    




\section{Discussion}

Many of our findings are unsurprising. Public health organizations favor familiar chart types, communicate primarily through temporal trends, generally follow established structural conventions, and still exhibit accessibility challenges. Together, however, they suggest something more interesting: \textit{the forces shaping visualization practice may operate less at the level of individual charts and more at the level of organizations.}

Looking at a single visualization naturally leads us to ask why a designer chose a line chart, omitted direct labels, or selected a particular color palette. Implicitly, we treat every visualization as an independent design decision. Looking across more than 4,000 visualizations tells a different story. The observed consistency in chart types, annotations, color usage, and organization-level design profiles suggests that many of these decisions are not made from scratch. Instead, they emerge from recurring organizational processes. This shift in perspective changes the unit of explanation. Rather than asking why this chart looks the way it does, it prompts us to examine why organizations repeatedly produce such charts. The answers are unlikely to be limited to visualization principles but also extend to communication goals, reporting workflows, templates, branding, governance, and institutional memory.

\subsection{Organizations Develop Stable Ways of Visualizing}
Viewed through this lens, our findings become less about individual design choices and more about stable organizational practices. Organizations appear to develop recurring visualization practices that persist across reports, topics, and authors.
%
%
%
%
%
%
The stability of design profiles suggests that organizations repeatedly solve similar communication problems using an established visual vocabulary. Likewise, recurring annotation practices and restrained color palettes likely reflect communication habits that have evolved over years of organizational practice rather than decisions made anew for every visualization. Branding requirements, inherited templates, software defaults, production timelines, and review processes all shape the final visualization. Similar questions can be asked of chart diversity, labeling conventions, and other design recommendations.


This perspective resonates with recent work on visualization style guides \cite{ottley2026consensus, choi2021toward}. Ottley et al. analyzed over 2,000 guidelines from across 52 organizations and found that guidelines are shaped by values such as clarity, efficiency, comprehensibility, and simplicity \cite{ottley2026consensus}. Choi et al. evaluated 226 guidelines and found that guidelines are commonly focused on problem-solving and enhancement \cite{choi2021toward}. These works further elaborate that practice can be seen as a product of an organization's values and culture. Furthermore, style guides explicitly document how organizations believe visualizations should be created, while the present work examines what organizations actually publish. Together, they suggest that organizations develop stable ways of visualizing that are both codified and enacted through everyday practice. Understanding how these practices emerge, evolve, and sometimes diverge from published guidance remains an important direction for future work.

\subsection{Closing the Loop Between Research and Practice}

Visualization research has traditionally asked how to design better charts. Our findings suggest an equally important question: How do better charts become standard practice? Understanding that process may be a grand challenge for empirical visualization research.

Evaluating departures from best practices is an important investigation in visualization research. Accessibility issues, for example, can be interpreted as poor design decisions. At the organizational scale, they instead point to a different question: \textit{Why do accessibility recommendations fail to become routine practice?} The answer is likely more complex than individual designer expertise. Joyner et al. conducted an in-the-wild analysis of visualization accessibility and interviewed over 100 practitioners \cite{joyner2022visualization}. Their findings identified three levels of uncertainty in the designer's evaluation of a visualization's accessibility: personal, operational, and organizational. The work further illuminates the complexity of visualization practitioners' decisions beyond skill level, chosen topic, or specific data task.


\section{Data Availability and Ethical Considerations}

The corpus comprises visualizations collected from 26 public health organizations, the majority of which are U.S. government agencies whose visualizations are in the public domain. Data collection was limited to publicly available web pages requiring no authentication, and organizations with detectable security restrictions (e.g., Cloudflare) were excluded from the corpus. Accordingly, we will release the complete set of public-domain visualizations, along with the derived annotations, metadata, codebook, analysis scripts, and archived source URLs, creating a reusable resource for visualization research. Visualizations from nonprofit and international organizations will not be redistributed because they may remain the intellectual property of their respective organizations; in these cases, we provide metadata, features, and archived source URLs to facilitate retrieval from the original sources where permitted. Researchers using the corpus for downstream applications, including computational model development, should ensure that their use complies with applicable copyright, licensing, and website terms. This study did not involve human participants and therefore did not require Institutional Review Board (IRB) approval.


\section{Limitations and Future Work}

Our work has several limitations. First, while the corpus is large and ecologically grounded, it is restricted primarily to public-facing visualizations from U.S. public health organizations and does not capture many international, non-English, internal, or interactive visualizations. Second, the codebook was initially developed by a single visualization expert before iterative refinement, introducing the possibility of subjective bias in the coding framework. Third, although our automated extraction and annotation pipeline was manually validated, it may still contain classification errors. Our conservative filtering criteria may have excluded some valid visualizations, while automated chart classification assigned a single primary chart family and may not fully represent hybrid or multi-view visualizations. Finally, because no manual review was performed on every collected image, a small number of non-visualization images may remain in the corpus.

Future work should extend this approach beyond public health to additional domains and examine how organizational visualization practices evolve over time, particularly during major public health events or policy shifts. An equally important direction is to connect corpus-level patterns with audience studies, to examine how visualization practices influence comprehension, trust, and decision-making. 

\section{Conclusion}

We presented a large-scale empirical characterization of real-world visualization practice through the lens of public health communication. By constructing and analyzing a corpus of more than 4,000 visualizations collected from over two dozen U.S. and international organizations, we revealed a design space that is highly concentrated, structurally uneven, and institutionally shaped. Our findings show that a small set of familiar chart forms dominates public-facing communication, important accessibility and interpretability risks remain widespread, and organizations exhibit stable design profiles shaped by communication norms and reporting goals.
Beyond the domain of public health, this work highlights the value of corpus-scale analysis as a methodological approach for visualization research.

\acknowledgments{%
	This work is supported in part by the National Science Foundation under Grants Nos. 2142977 and 2330245, which support the Engineering Research Center for Carbon Utilization Redesign through Biomanufacturing-Empowered Decarbonization (CURB). M. Hines was supported by the following fellowship programs in sequential order: Clare Boothe Luce and NRT-AI: AI Advancements and Convergence in Computational, Environmental, and Social Sciences (AI-ACCESS) (No.  2244165). %
}

\bibliographystyle{abbrv-doi-hyperref}

\bibliography{template}

\clearpage
\onecolumn
\appendix 
\crefalias{section}{appendix} 

\section{Sources That We Selected Public Health Organizations}

\label{appendix:orgs_sources}
\begin{table*}[h]
    \caption{The complete list of sources (with their URLs) we used to select public health organizations for our study. }
    \label{tab:codebook}
    \scriptsize
    \centering 
    \begin{tabular}{|C{5cm}|C{10cm}|}
        \hline
        \textbf{Source} & \textbf{URL} \\
        \hline
        2025 Forbes Top Charities & \url{https://www.forbes.com/lists/top-charities/} \\
        \hline
        NIH Institutes and Centers & \url{https://www.nih.gov/institutes-nih/list-institutes-centers} \\ 
        \hline 
        HHS Agencies and Offices & \url{https://www.hhs.gov/about/agencies/hhs-agencies-and-offices/index.html} \\
        \hline 
        UN and Multilateral Organizations & \url{https://www.un.org/en/about-us/un-system} \\
        \hline 
        University of Illinois Chicago, School of Public Health & \url{https://publichealth.uic.edu/global-health-program/applied-practice-experience/un-multilateral-organizations/} \\
        \hline
    \end{tabular}
\end{table*}

\section{Finalized Codebook}
\label{appendix:codebook}
\renewcommand{\arraystretch}{1.5}
\begin{table*}[h!]
    \caption{Description of the finalized codebook with 6 major categories and the variables nested in each. Data types of the variables are in parentheses. \textit{Assignment} refers to how the variable was designated.}
    \label{tab:codebook}
    \scriptsize
    \centering 
    \begin{tabular}{|C{2cm}|Y{2cm}|Y{7cm}|C{2cm}|}
        \hline
        \textbf{Theme} & \textbf{Short Description} & \textbf{Variables (with Data Type)} & \textbf{Assignment} \\
        \hline
        Chart Family & High-level chart type & Chart Type (Single Option): Bar Chart, Area Chart, Line Chart, Scatter Plot, Map, Table, Pie Chart, Histogram, Box Plot, Heatmap, Flow Diagram, Network Variable, Tree Diagram, Timeline, Dashboard, Infographic, Other & GPT \\
        \hline
        Data Context & Data task & Data Context (Single Option): Temporal Trend, Geographic, Categorical Comparison, Distribution, Composition Part to Whole, Ranking Ordered, Correlation Relationship, Flow Process, Hierarchical, Network Relational, Multivariate, Text Narrative, Mixed & GPT \\
        \hline
        Structural Features & Basic structure elements & Direct Labels (Boolean), Legend (Boolean), Grid Lines (Boolean), Title (Boolean), Y-Axis (Boolean), X-Axis (Boolean), Labeled Axes (Boolean), Multiple Panels (Boolean), Missing Data (Boolean) & GPT \\
        \hline
        Visual Embellishments & Decorative elements & Iconography (Boolean), Photograph (Boolean), Medical Imaging (Boolean), Unique Shapes or Lines (Boolean), Text Annotations (Boolean), X-Axis (Boolean), Text Annotations (Boolean) & GPT \\
        \hline
        Color & Color palette & Color Blind Risky (Boolean), Grayscale (Boolean), Palette Size (Number), HTML HEX Codes (List), HSV Color Names (List) & GPT, Measurement \\
        \hline
        Text & Text inside the chart & Visualization Text (Text), Text Density (Number) & OCR, Measurement \\
        \hline
    \end{tabular}
\end{table*}

\section{Preliminary Meta Data Analysis}
\label{appendix:meta_data}

Along with the scraping of visualizations, we collected web metadata fields including Last-Modified Date, Title, Author, and Description. We considered analyzing the timeline of visualizations produced and performing topic modeling. In this section, we discuss our decisions and preliminary findings. We encourage future work to explore the metadata further in their analyses.

\subsection{Timeline}
The Last-Modified date was collected from the HTTP header response of the webpage for the scraped image. The Last-Modified date may not be a reliable metric given the development of the visualizations in our corpus. Approximately 50\% of the visualizations (n = 2,207) have a Last-Modified date. A webpage with dynamic content, e.g., scripts executed on request, can update the Last-Modified date to the date the visualization was collected (n = 426). The Last-Modified date may not reflect the actual timeline when a visualization was created; we have observed that organizations reuse many visualizations, as further demonstrated by the removal of duplicate images in Section 3.4. We did not proceed with a timeline analysis of the collected visualizations; however, for completeness, we retained the Last-Modified date field in the public dataset.

\subsection{Topic Modeling}
We collected Title, Author, and Description from the meta tags of the HTML content related to the webpage of a visualization. Description and Author are applied unevenly across the visualization corpus: approximately 73\% of visualizations have descriptions (n = 3,122), and less than 2\% have a specified author (n = 65). All visualization webpages have titles; therefore, we continued with topic modeling of the title field to ensure an even representation of the corpus. We had difficulty finding meaningful insights from the topic groups, which we suspect is due to two potential reasons. 1) Titles can be pretty short, so topic models can easily group items on trivial words. It would take fine-tuning and editing of the titles to potentially garner meaningful insights. 2) Many of the public health organizations in our study have a specific purpose or cause, such as cancer, smoking, and mental health. Therefore, in our preliminary analysis, we identified topic groups based on the organizations' specific cause. For instance, smoking, drugs, and drinking were all one topic group. Ultimately, we found these groupings inconsequential given our primary focus on overall design patterns in the public health space.

\section{Examples of palette accessibility under simulated color vision deficiency}
\label{appendix:color_blind}

\begin{figure*}[!h]
    \centering
    \includegraphics[width=\linewidth, alt={Figure 5 consists of a table-like format where a visualization is presented on the far-left column and other columns describe if it's considered color-blind risky in consideration of particular CVD conditions.}]{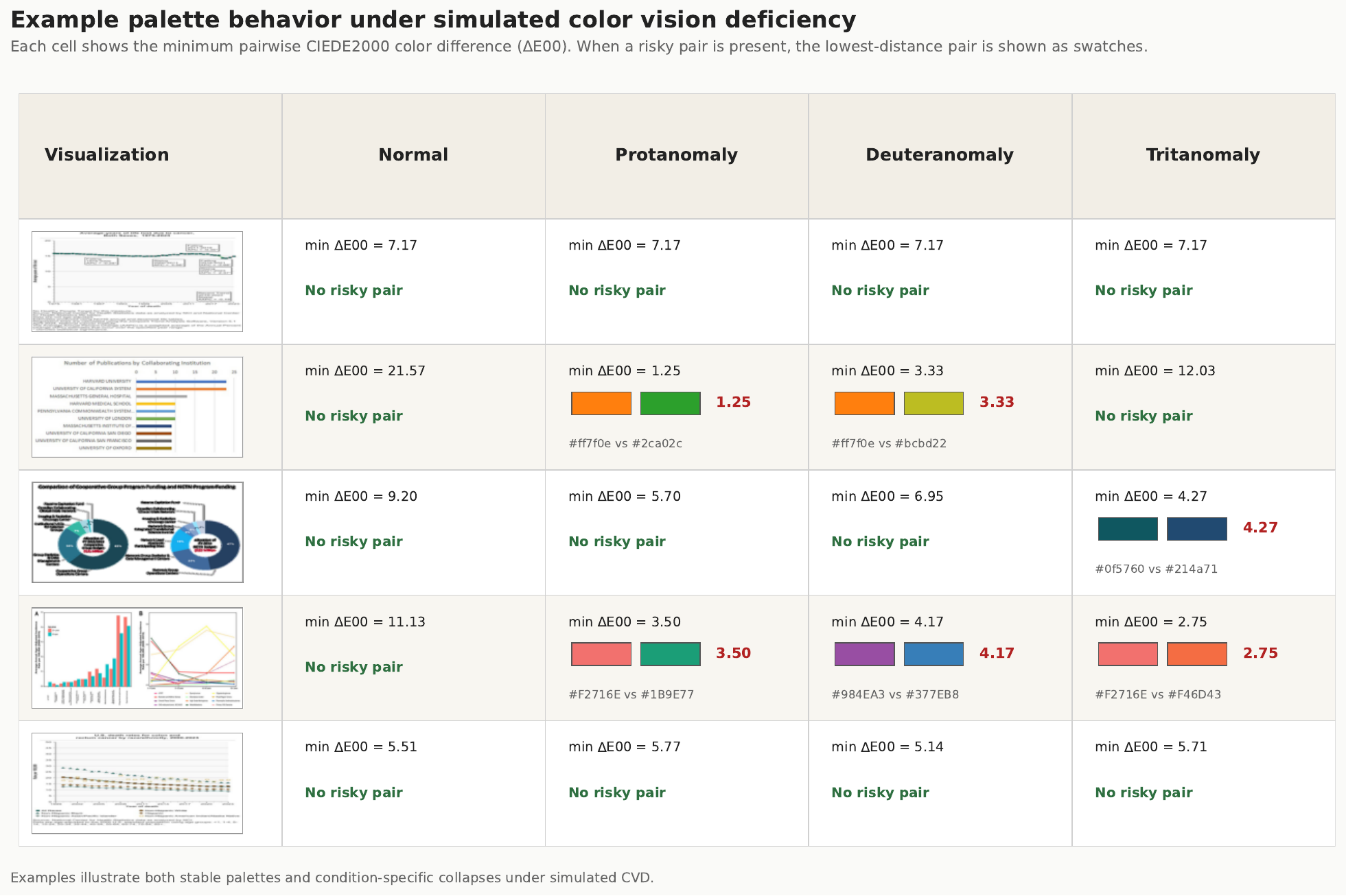}
    \caption{Representative examples from the corpus illustrating palette behavior under normal vision and three simulated color vision deficiency (CVD) conditions: protanomaly, deuteranomaly, and tritanomaly. Each cell reports the minimum pairwise perceptual color difference using CIEDE2000 ($\Delta E_{00}$). When a risky color pair is detected, the lowest-distance pair is shown as color swatches and annotated with its corresponding $\Delta E_{00}$ value. These examples illustrate both stable palettes and condition-specific color collapses that may reduce distinguishability under simulated CVD.}
    \label{fig:color_accessibility}
\end{figure*}

\clearpage 

\section{Visualization Examples from Design Profiles}
\label{appendix:vis_examples}
\begin{figure*}[!h]
    \centering
    \includegraphics[width=\linewidth, height=0.9\textheight, alt={Figure 6. showcases multiple visualizations from the corpus and they're grouped by design profiles.}, keepaspectratio]{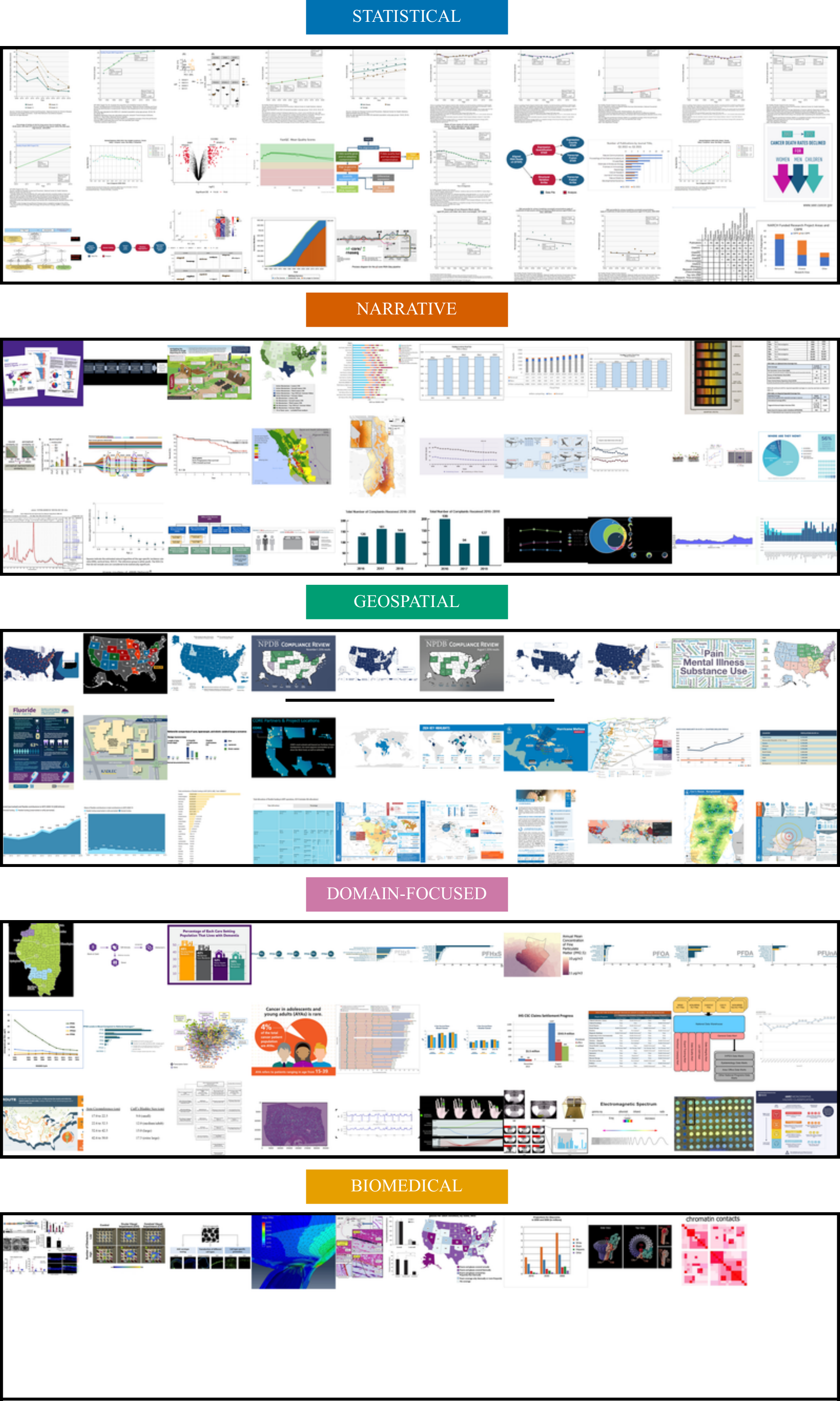}
    \caption{
   Sampled visualizations of all five design profiles.
    }
    \label{fig:profile_examples}
\end{figure*}







\end{document}